\documentclass[pdflatex,sn-mathphys-num]{sn-jnl}

\usepackage{graphicx}%
\usepackage{multirow}%
\usepackage{amsmath,amssymb,amsfonts}%
\usepackage{amsthm}%
\usepackage{mathrsfs}%
\usepackage[title]{appendix}%
\usepackage{xcolor}%
\usepackage{textcomp}%
\usepackage{manyfoot}%
\usepackage{booktabs}%
\usepackage{algorithm}%
\usepackage{algorithmicx}%
\usepackage{algpseudocode}%
\usepackage{listings}%
\usepackage{soul}%
\usepackage{braket}%
\usepackage{bm}%

\definecolor{DarkGreen}{RGB}{20,120,70}

\theoremstyle{thmstyleone}%
\theoremstyle{thmstyletwo}%

\theoremstyle{thmstylethree}%

\begin{document}

\title[Article Title]{Generalized thermodynamic closure in ultrafast phonon dynamics}


\author[1]{\fnm{Sheng} \sur{Qu}} 
\equalcont{These authors contributed equally to this work.}

\author[2]{\fnm{Jiyong} \sur{Kim}}
\equalcont{These authors contributed equally to this work.}

\author[1,3]{\fnm{Jaco J.} \sur{Geuchies}}

\author[4]{\fnm{Sergey} \sur{Kovalev}}
\author[4]{\fnm{Jan-Christoph} \sur{Deinert}}
\author[4]{\fnm{Thales} \sur{de Oliveira}}
\author[4]{\fnm{Alexey} \sur{Ponomaryov}}
\author[4]{\fnm{Min} \sur{Chen}}
\author[4]{\fnm{Nilesh} \sur{Awari}}
\author[4]{\fnm{Igor} \sur{Ilyakov}}

\author[1]{\fnm{Mischa} \sur{Bonn}}

\author*[1,2]{\fnm{Heejae} \sur{Kim}}\email{heejaekim@postech.ac.kr}

\affil*[1]{\orgdiv{Department of Molecular Spectroscopy}, \orgname{Max-Planck Institute for
Polymer Research}, \orgaddress{\city{Mainz}, \postcode{55128}, \country{Germany}}}

\affil[2]{\orgdiv{Department of Physics}, \orgname{Pohang University of Science and Technology}, \orgaddress{ \city{Pohang}, \postcode{37673},  \country{Korea}}}

\affil[3]{\orgdiv{Leiden Institute of Chemistry}, \orgname{Leiden University}, \orgaddress{\city{Leiden}, \postcode{2333CC}, \country{Netherlands}}}

\affil[4]{\orgdiv{Institute of Radiation Physics}, \orgname{Helmholtz-Zentrum Dresden-Rossendorf}, \orgaddress{\city{Dresden}, \postcode{01328},  \country{Germany}}}


\abstract{Driven-dissipative dynamics underlie a wide range of nonequilibrium phenomena in quantum materials, yet reduced descriptions beyond the quasi-equilibrium picture remain difficult to establish. Here, we experimentally demonstrate that a resonantly driven phonon mode admits a generalized thermodynamic description in which coherence and energy jointly organize the nonequilibrium evolution. Beyond a threshold driving field strength, we observe a delayed ultrafast response of a coherently driven phonon mode. Combined with experimentally constrained Lindblad dynamics, we show that this delay reflects the finite-time spreading of excitations across many phonon levels. At the same time, the full density-matrix trajectories for three driving conditions collapse onto a common surface defined by energy and coherence. Our results establish a coherence-extended thermodynamic regime for driven phonons and provide a framework for broader state engineering in driven-dissipative bosonic excitations.}





\maketitle
\section{Introduction}\label{sec1}

Driven-dissipative quantum systems lie at the intersection of quantum information science~\cite{Verstraete2009DissipationQI}, Floquet engineering~\cite{Eckardt2017FloquetReview}, and active matter~\cite{Marchetti2013ActiveMatter}. A central challenge in studying these systems is identifying a reduced set of state variables that captures the essential physics without resolving the full microscopic complexity. Whereas equilibrium statistical mechanics provides a general answer through thermodynamic closure, no comparably universal principle is known far from equilibrum~\cite{Rodrigues2024CoherenceThermo}. In equilibrium, conservation laws and the thermodynamic state postulate guarantee that a small number of macroscopic variables (e.g.\ $P, T$) uniquely determine the state and its entropy~\cite{Callen1985Thermodynamics}. Close to equilibrium, this framework remains robust, and formalisms such as Mori--Zwanzig projection rigorously separate slow collective dynamics from fast microscopic noise~\cite{Mori1965Projection,Zwanzig1961Memory}. Far from equilibrium, however, driving and dissipation can undermine these safeguards by violating fundamental conservation laws, thereby obscuring the boundary between emergent macroscopic dynamics and microscopic fluctuations. In such regimes, the system's evolution may no longer be captured by a single-valued thermodynamic manifold~\cite{Rodrigues2024CoherenceThermo}.

From an experimental standpoint, ultrafast spectroscopy is uniquely positioned to resolve these evolutionary paths with femtosecond precision. By synchronizing non-adiabatic excitation with real-time observation, these techniques offer a window into the transient trajectories of far-from-equilibrium matter~\cite{deLaTorre2021Nonthermal,Mitrano2020Probing}. Despite this potential, experimental analysis has conventionally defaulted to quasi-equilibrium variables---such as effective temperatures~\cite{Allen1987Thermal,Caruso2022Ultrafast}---or highly truncated few-level models~\cite{Preuss2022Resonant, kim2017direct}. Such reduced-order descriptions implicitly assume that the system's physics can be confined to a narrow subset of its Hilbert space. However, these approximations are fundamentally ill-equipped to describe regimes where the drive forces the system into a higher-dimensional state space. Consequently, a comprehensive experimental characterization of how these trajectories escape and eventually return to the thermodynamic manifold has remained elusive~\cite{deLaTorre2021Nonthermal,Maiuri2020Ultrafast}.

Here, we experimentally demonstrate the breakdown of quasi-equilibrium thermodynamic closure in a driven-dissipative many-body system. By resonantly driving a specific lattice vibration in a crystalline solid with intense terahertz (THz) pulses, we interrogate the electronic evolution with sub-picosecond resolution. Above a critical threshold, we observe a delayed electronic response that cannot be explained by standard few-level truncations. Through simulations using an experimentally constrained Lindblad master equation, we reveal that this delay signifies the delayed multi-level occupation spreading. Yet, the full density-matrix evolution remains confined to a low-dimensional manifold: for three driving conditions, the trajectories collapse onto a common surface $S=S(E,C)$, with E the phonon energy and C the total coherence. Thus, our results provide evidence that coherently driven lattice excitations can realize a coherence-extended thermodynamic regime in solids.

\begin{figure}[h]
\centering
\includegraphics[width=0.9\textwidth]{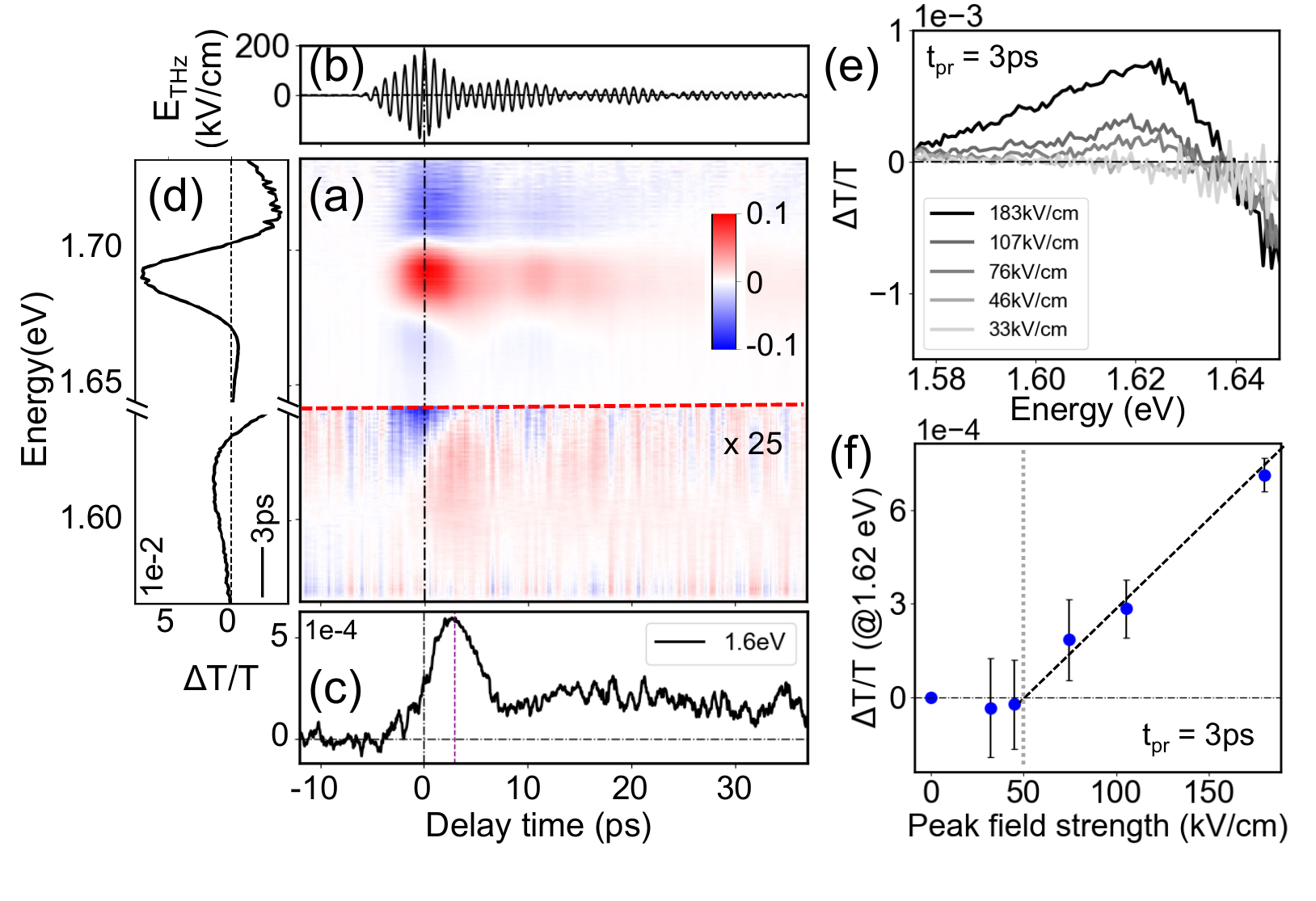}
\caption{Experimental realization of a driven-dissipative phonon system far from equilibrium.  (a) The differential transmission spectrum, $\Delta T /T$, of MAPI at 100 K via the THz pump pulse centered at 1 THz. The spectra at the probe energy below 1.65 eV are magnified to better display the sub-band-gap response (see main text). (b) The temporal profile of the pump THz pulse. (c) The dynamics of the differential transmission at the probe energy of 1.6 eV. The 1.6 eV response reaches its maximum after $\sim 3$ ps from the maximum of the THz pump. (d) The differential transmission spectrum at the pump-probe delay time, $t_{\rm pr}$, of 3 ps. (e) The transient spectra of the suppressed sub-gap transition at $t_{\rm pr}=$ 3 ps at various peak field strengths of the drive THz pulses. (f) The threshold behavior of the suppression of sub-gap transition upon the drive peak field strengths.}\label{fig1} 
\end{figure}

\section{Results}\label{sec2}
To achieve a genuinely far-from-equilibrium state of matter, we use intense narrowband THz pulses from the accelerator-based TELBE THz facility \cite {green2016high, kovalev2017probing}, where the spectral brightness is three orders of magnitude higher than that of typical $\mathrm{LiNbO_3}$-based (100 kV/cm level) THz tabletop sources \cite{kim2017direct}. The frequency is tuned to be resonant (1 THz) with, slightly detuned (0.8 THz) from, and non-resonant (0.3 THz) with the target phonon mode (centered at 1 THz) of methylammonium lead iodide perovskite (($\mathrm{CH_3NH_3PbI_3}$, MAPI, see Supplementary Fig. S1 (a-b)). Upon the strong, transient drive, the subsequent electronic response from the near-gap transitions (Supplementary Fig. S1 (c)) that are strongly coupled with the particular phonon mode \cite{qu2025mode, kim2017direct} is captured by subsequent near-infrared probe pulses. 

The thus obtained electronic response via a driven collective lattice mode (Fig. 1(a)) is displayed together with the temporal profile of the THz driving pulse shown above (on the same time axis, Fig. 1(b)). At the experimental temperature (100 K), the system is in its orthorhombic phase ($T_c \sim$160 K) where the band gap is located around 1.67 eV \cite{milot2015temperature}. A noticeable, instantaneous response in the probe energy of 1.66-1.72 eV is the expected Franz-Keldysh effect \cite{aspnes1974band, schmitt1989linear, kim2017direct}. It has a purely electronic origin arising from the electric field-modified Bloch-like electron wave function \cite{kim2017direct, berghoff2021low}.

Unexpectedly, above a certain drive threshold ($>$ peak field strength of 50 kV/cm with the spectral width of 0.2 THz), a distinct electronic response in the probe energy of 1.58-1.63 eV ($\sim 60$ meV below the band gap) starts to appear and peaks approximately 3 ps after the driving pulse (Fig 1. (c-f)). The positive differential transmission (Fig 1. (d)) indicates a suppression of the sub-gap absorption, which corresponds to the significant tail below the band gap (Supplementary Fig. S1 (c)) due to the dynamic structural disorder \cite{bechtel2018octahedral}. Such a dynamically delayed, suppression of the sub-gap absorption (i) emerges only when the THz pump frequency is (near-) resonant (i.e. 0.8 and 1 THz), not when non-resonant (i.e. 0.3 THz,  Supplementary Fig. S2) (ii) exhibits identical delay times when the THz field strength varies (with the Fig 1(a) obtained by the maximum, 183 kV/cm, Supplementary Fig. S3(a-b)), and (iii) persists over the temperature range of 100-140 K (Supplementary Fig. S3(c-e)). As the frequency selectivity of the THz driving pulse testifies to, this distinct electronic response directly reflects the evolution of a strongly driven system via a specific collective lattice mode above a certain drive threshold. 

Before we further investigate the evolution of this driven lattice system, we briefly pause to rule out other possibilities that might have provided an alternative interpretation of the observed electronic response:  (i) the average temperature rise, estimated to be $<\sim5$ K (See SI, Suppl. Fig. S4), by the total accumulated energy from the pulse trains (85 mW at 1 THz) is expected to show negative differential transmission (Fig 2(a)). (ii) THz-induced structural phase transition, if symmetry-allowed, is also expected to show negative differential transmission, due to the spectral transfer into the near gap of the high temperature (tetragonal) phase (above 1.55 eV) from that of the orthorhombic phase (above 1.65 eV) (Fig 2(b)). (iii) In the case of THz-induced electronic interband transition, a positive differential transmission due to bleaching is expected, but only in the above band gap range, not the observed below band gap range. Furthermore, the dynamic response predicted from optical Bloch equations \cite{haug2009quantum} based on the experimental THz field profile barely decays within the experimental window (Fig 2(c), see Methods), which shows a stark contrast to our observed, clear relaxation within 10 ps (Fig 1(c)). (iv) Franz-Keldysh effect from possibly co-existent tetragonal phase below $T_c$ \cite{schotz2020investigating,kong2015characterization} shows a distinctive spectral shape (dominantly tetragonal phase 160 K, Fig. 2(d)) from the observed transient spectrum (100 K, Fig. 2(d)). Therefore, we can assure that we achieved a strongly driven lattice system, monitored through an electronic lens, i.e., the suppression of the sub-gap transitions.

\begin{figure}[h]
\centering
\includegraphics[width=0.9\textwidth]{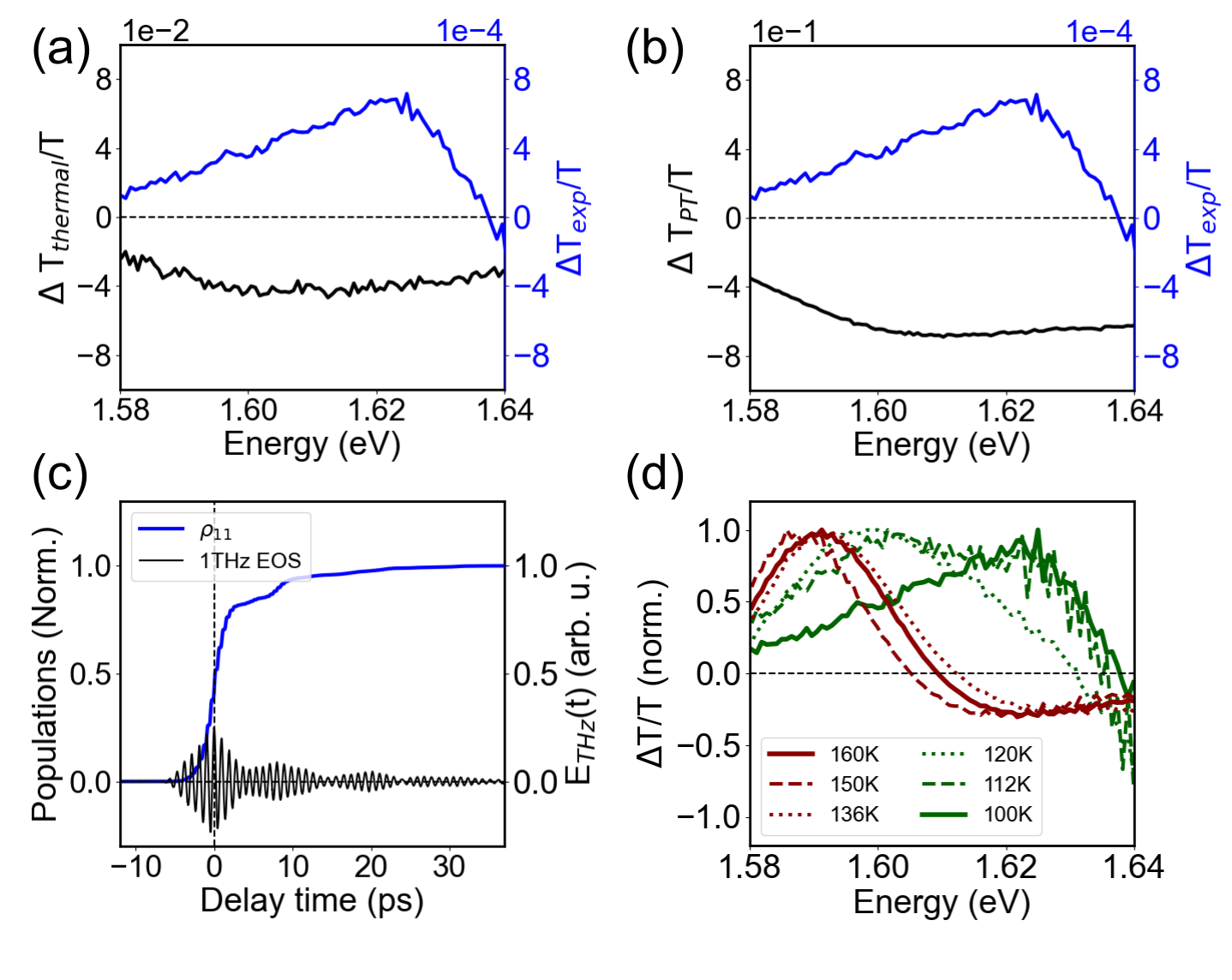}
\caption{Failure of alternative interpretations for the observed electronic response. (a) (a, b) The expected differential transmission spectra (black) obtained from the temperature-dependent, steady-state absorption curves (Fig. S1(c)). $\Delta T_{\rm thermal}/T$ ($\Delta T_{\rm PT}/T$)is estimated from $T_{120K}-T_{100K}$ ($T_{180K}-T_{100K}$), respectively. Each of the estimated curves is compare with the pump-probe differential spectrum $\Delta T_{\rm exp}/T$ (blue) shown in Fig. 1(e). (c) The expected dynamic response (blue) of THz-induced electronic interband transition, calculated from optical Bloch equations (Methods) based on the experimental THz field profile (black). (d) Temperature dependence of the differential transmission spectra. Burgundy curves are dominated by the Franz Keldysh effect of coexisting tetragonal crystallites near the critical temperature (136 to 160 K). Solid and dashed green curves (100 and 112 K) show no discernible Franz Keldysh effect, with a distinctive spectral feature. The dotted green curve (120 K) shows a mixture of Franz Keldysh effect and purely phonon-induced suppression of sub-gap transitions (see text). } \label{fig2}   
\end{figure}

\begin{figure}[h]
\centering
\includegraphics[width=0.9\textwidth]{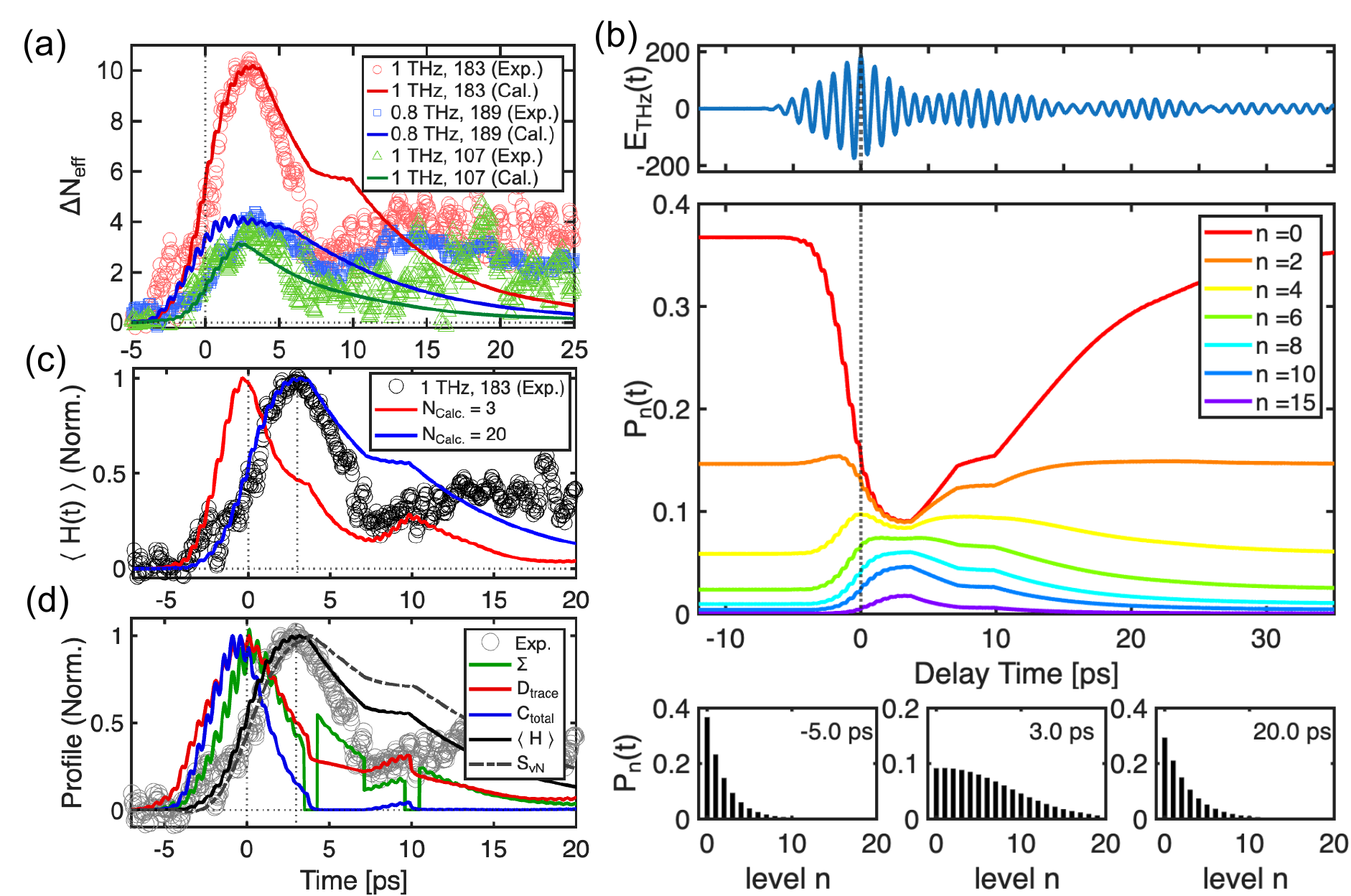}
\caption{Modeling the driven phonon as an open quantum system. (a) Effective participation number ($N_{\rm eff}(t)$) from the experimentally constrained Lindblad simulation (solid curves) compared with the measured responses (open circles) for three drive conditions: 1 THz, 183 kV/cm (red); 0.8 THz, 189 kV/cm (blue); and 1 THz, 107 kV/cm (green) (b) Simulated population dynamics of individual phonon levels (middle) and population distributions at selected pump-probe delays (bottom; -5, 3, and 20 ps) for the 1 THz, 183 kV/cm drive condition; the corresponding experimental THz waveform is shown above (top). (c) Comparison of simulations with Hilbert-space truncation to $N=$ 3 (red) and $N=$20 (blue), together with the measured dynamics (black circles) (d) Comparison of scalar observables extracted from the simulation:  entropy production ($\Sigma$, green), trace distance from the equal-energy Gibbs state ($D_{\rm trace}$, red), total coherence ($C_{\rm total}$, blue), energy expectation value ($\langle H \rangle$, black solid), and von Neumann entropy ($S_{\rm vN}$, dark gray dashed), together with the measured dynamics (gray circles).}\label{fig3}
\end{figure}

\section{Discussion}\label{sec3}

\subsection{Phonon Population Spreading: the Observable Proxy}

To explore the root of the delayed response, we compare three drive conditions above the threshold field of $\sim$ 50 kV/cm: (i) resonant pulse at 1 THz with a peak field strength of 183 kV/cm (ii) near-resonant pulse at 0.8 THz with comparable peak field strength of 189 kV/cm, and (iii) resonant pulse at 1 THz with a lower peak field strength of 107 kV/cm. In all three cases, the delayed maximum response persists, whereas the peak amplitude depends strongly on drive condition (Fig. 3(a)), even though the transient spectra remain nearly identical (Suppl. Fig. S3(d)). This delayed the maximum by about 3 ps is incompatible with an effective-temperature picture based on a three-level truncation  \cite{kim2017direct} (see below). We therefore model the driven phonon as an anharmonic open quantum system with access to higher-lying phonon states, by solving a Lindblad master equation ~\cite{Lindblad1976,Gorini1976,Breuer2002,Carmichael1999} using the measured THz waveforms and experimentally constrained material parameters~\cite{la2016phonon,Whalley2016MAPIAnharmonicity,Brivio2015MAPI} (Methods, Fig. 3(b)). The much larger peak response under resonant driving in condition (i) than under near-resonant driving in condition (ii), despite the higher pulse energy of the latter(Fig. 3(a)), shows that the dynamics are not governed by deposited energy alone, but by a nonequilibrium restructuring of the dissipative pathways under strong driving~\cite{deLaTorre2021Nonthermal}. Remarkably, a single set of experimentally constrained dissipative parameters reproduces the dynamics across all three drive conditions when combined with a field-dependent suppression of relaxation above $\sim$ 55 kV/cm (Methods).

To identify an experimentally relevant proxy for the measured transients, we evaluate scalar observables from the full instantaneous density matrix, $\rho(t)$ obtained from the Lindblad dynamics (Method). Among the quantities considered, two exhibit temporal profiles whose maxima lie closest to the experimental delay: the energy expectation value $\langle H(t) \rangle$ and the effective participation number,
\begin{equation}
    N_{\rm eff}(t)=\left(\sum_n P_n(t)^2\right)^{-1},
\qquad
P_n(t)=\rho_{nn}(t),
\end{equation}
which quantifies the effective number of phonon levels significantly occupied by the instantaneous population distribution~\cite{Bell1970,Wegner1980}. Whereas $\langle H(t) \rangle$ provides a macroscopic measure of the energy stored in the driven mode, $N_{\rm eff} (t)$ measures how broadly the population is distributed across the phonon energy levels~\cite{Breuer2002,Bell1970,Wegner1980}.
Crucially, although both $\langle H(t) \rangle$ and $N_{\rm eff} (t)$ reproduce the observed $\sim$ 3 ps delay, only $N_{\rm eff} (t)$ captures the relative signal amplitudes across the resonant and near-resonant drive conditions (Supplementary Fig. S5(a)). By contrast, $\langle H(t) \rangle$ systematically overestimates the near-resonant response, where energy is absorbed but the population remains relatively concentrated in the lower-lying phonon states (Fig. S6). This comparison suggests that the probe is sensitive not simply to the total absorbed energy, but to the extent of population spreading across phonon levels, or to a closely related nonlinear functional of the instantaneous population distribution.

We further find that a three-level truncation of the phonon Hilbert space is insufficient to account for the observed $\sim$ 3 ps delay(Fig 3(c)). In the full Lindblad calculation with $N=$20 levels, $N_{\rm eff}$ reaches a maximum value of $\sim$ 14.5 (Supplementary Fig. S5(b); Methods), demonstrating that the driven state spreads over a large number of phonon levels. In contrast, when the dynamics are artificially restricted to $N=$3, all scalar observables peak close to the maximum of the THz pulse. The delayed response therefore emerges as an intrinsically many-level phenomenon, beyond the reach of any few-level description.

\begin{figure}[h]
\centering
\includegraphics[width=0.9\textwidth]{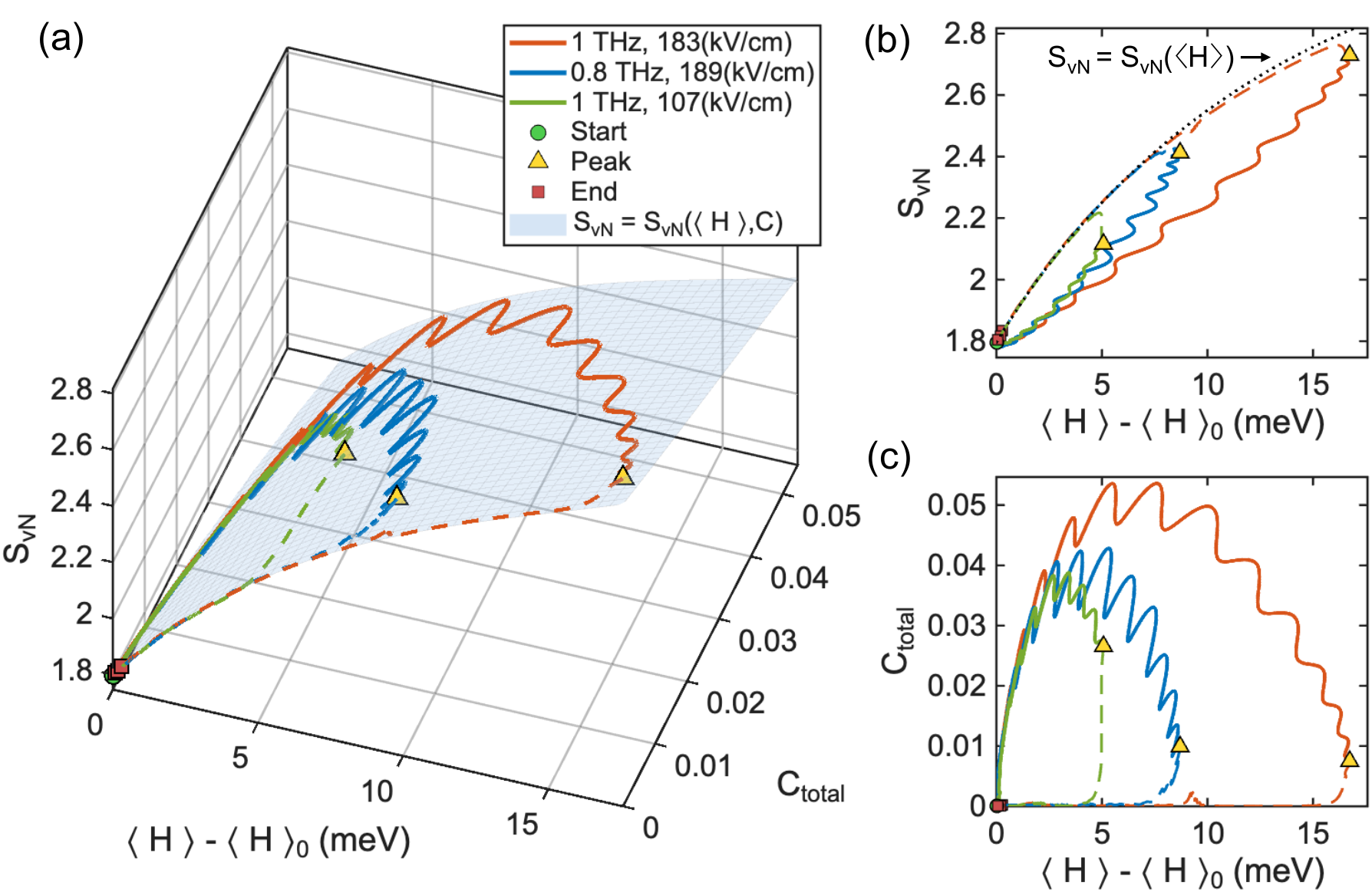}
\caption{Coherence-extended thermodynamic closure. (a) Three-dimensional representation of the phonon energy expectation value ($\langle H \rangle$), the total coherence ($C_{\rm total}$), and the von Neumann entropy ($S_{\rm vN}$), extracted from the full density matrix ($\rho(t)$) obtained from the experimentally constrained Lindblad simulation. The coherence-extended thermodynamic surface, $S_{\rm vN}= S_{\rm vN}(\langle H \rangle, C_{\rm total})$, is shown as a semi-transparent light-blue layer. (b) Hysteresis in the energy-entropy plane, shown together with the equilibrium thermodynamic relation, $S_{\rm vN}=S_{\rm vN}(\langle H \rangle)$, obtained from the equal-energy Gibbs reference state (dotted black). (c) Hysteresis in the energy-coherence plane. In all panels, solid curves denote the excitation trajectories up to the maximum of the energy (yellow triangle) and dashed curves the subsequent relaxation trajectories for the three experimental drive conditions (red: 1 THz with 183 kV/cm, blue: 0.8 THz with 189 kV/cm, green: 1 THz with 107 kV/cm).}\label{fig4}
\end{figure}


\subsection{Coherence-extended Thermodynamic Closure} 

The remaining observables extracted from the full density matrix are the distance from thermal equilibrium at the same energy ($D_{\rm trace}(t)$), the total coherence ($C_{\rm total}(t)$), the von Neumann entropy ($S_{\rm vN}(t)$), and the entropy production rate ($\Sigma(t)$) (Fig 3(d)). To quantify the instantaneous departure from equilibrium, we compare $\rho(t)$ with the Gibbs state $\rho_{\rm G}(t)$ having the same mean energy, determined by solving $\rm Tr(H_0 \rho_{\rm G}(t))= \rm Tr(H_0 \rho(t))$ \cite{Jaynes1957,Callen1985Thermodynamics}. The nonequilibrium distance is then measured by the trace distance, $D_{\rm trace}(\rho(t), \rho_{\rm G}(t))=\frac{1}{2} \rm{Tr} \sqrt{(\rho(t)-\rho_{\rm G}(t))^{\dagger}(\rho(t)-\rho_{\rm G}(t))}$ \cite{NielsenChuang2000}. We further define the total quantum coherence as, $C_{\rm total}(t)= \sum_{n \neq m}|\rho_{nm}(t)|^2$, which quantifies the total off-diagonal weight of the density matrix in the phonon energy eigenbasis and therefore the basis-dependent quantum coherence generated during the dynamics \cite{Baumgratz2014}. The von Neumann entropy, $S_{\rm vN}(t) = -\rm Tr(\rho(t) ln \rho(t))$, measures the mixedness of the reduced state \cite{NielsenChuang2000}. Finally, the entropy production rate, $\Sigma(t)$, evaluated using the Spohn formalism \cite{Spohn1978}(Methods), provides a measure of the irreversible contribution to the driven open-system dynamics \cite{Spohn1978,Breuer2002}. 

As shown in Fig 3(d), the scalar observables reveal a clear temporal hierarchy that resolves the ultrafast phonon dynamics into distinct physical stages. Quantities most directly associated with the action of the drive - including the total coherence, the trace distance to the equal-energy Gibbs state, and the entropy production rate - peak close to the maximum of the THz field, indicating the prompt generation of a coherent nonequilibrium state. By contrast, the energy expectation value and the effective participation number peak later, coincident with the experimental response (Fig 3(d)). This identifies the measured delay not with the instant of maximal driving, but with the time at which the phonon population reaches its widest effective distribution over the accessible levels. The slightly later maximum of the von Neumann entropy shows that dissipative mixing continues to develop during and beyond this population-spreading stage. The dynamics therefore proceeds through a sequence of coherent excitation, delayed multilevel population spreading, and subsequent maximal mixedness. This interpretation is further reinforced by the comparison with reduced-Hilbert-space calculations, which fail to reproduce the delayed maximum (Fig. 3(b)). Altogether, these results demonstrate that the experimentally observed delay reflects the finite-time buildup of population over an extended phonon ladder rather than the instantaneous action of the drive.

Finally, we investigate whether the far-from-equilibrium driven dynamics admit a reduced thermodynamic closure in terms of a small set of scalar observables. Remarkably, for all three experimental drive conditions, the density-matrix trajectories collapse onto a common surface $(S_{\rm vN}=S_{\rm vN}(\langle H\rangle,C_{\rm total})$; Fig. 4). This collapse shows that, although the microscopic state evolves in a high-dimensional Hilbert space, its entropy is nearly fixed once the energy and coherence are specified. By contrast, the conventional effective-temperature picture, in which the equilibrium relation ($S=S(\langle H \rangle)$) defines a sufficient thermodynamic manifold \cite{Allen1987Thermal,Caruso2022Ultrafast}, clearly breaks down, as evidenced by the hysteresis in the energy-entropy plane (Fig. 4(b)).  The coherence resolves states with similar energy but distinct quantum character (Fig. 4(c)), indicating that energy alone is insufficient to parametrize the nonequilibrium evolution. The surface $S_{\rm vN}=S_{\rm vN}(\langle H \rangle,C_{\rm total})$ therefore provides an explicit realization of a coherence-extended thermodynamic description in a crystalline solid. The fact that varying the drive strength and waveform moves the system along the same reduced geometry further suggests that this description is robust across a range of nonequilibrium driving conditions. In this picture, the agreement between $N_{\rm eff}(t)$ and the measured transient is naturally interpreted as the observable manifestation of motion along this coherence-extended nonequilibrium manifold.

\section{Conclusion}\label{sec4}

In conclusion, we have experimentally realized and quantiatively characterized a far-from-equilibrium driven phonon state in which the measured response tracks the population spreading across phonon levels rather than the instantaneous drive amplitude or the total absorbed energy. The pronounced delay of the response following the THz-field maximum cannot be captured by either standard few-level truncations or quasi-equilibrium descriptions. Instead, the full density-matrix dynamics reveal a clear temporal sequence: coherent excitation is followed by delayed multi-level population spreading and only then by maximal entropy growth. This decoupling generates a pronounced hysteresis in the energy-entropy plane, indicating that the driven evolution departs from the conventional thermodynamic manifold before relaxation returns toward it. At the same time, the underlying state trajectory remains confined to a reduced nonequilibrium surface ($S_{\rm vN}=S_{\rm vN}(\langle H \rangle,C_{\rm total})$), demonstrating that energy alone is insufficient to specify the macroscopic state, and that coherence must be retained as an independent nonequilibrium coordinate. Our findings therefore provide evidence for a coherence-extended thermodynamic description of coherently driven lattice excitations in a crystalline solid, beyond the effective-temperature paradigm. More broadly, because coherent driving, dissipation and multilevel structure are generic to collective modes, this framework should extend beyond the present material and phonon mode, opening up the possibility of engineering open quantum systems with improved reduced descriptions, more refined control targets, and more faithful state characterization.

\section{Methods}
\subsection{Experimental Details}
A high-repetition-rate super-radiant THz source (repetition rate = 50 kHz; peak electric field ${\sim}$190 kV/cm; FWHM = ${\sim}$0.1 THz) developed in a Terahertz facility at ELBE (TELBE) is used as pump beam and synchronized to an external Ti:Sapphire laser system (center wavelength = 800 nm, pulse duration = 35 fs, repetition rate = 100 kHz) \cite{green2016high, kovalev2017probing}. The center frequency of this terahertz source is tunable. In this experiment, we used center frequencies of 0.3, 0.8, and 1 THz pump pulses to excite the sample. A THz fundamental band pass filter is set at the beginning of the THz beam path to purify the frequency components in the THz pulses. The time profile of the pump THz pulses is characterized by the standard electro-optical sampling technique via a ZnTe (110) crystal \cite{reimann2007table}. The white-light supercontinuum probe beam is generated by focusing the 800 nm fs-pulse into a 5mm-thick sapphire crystal, and a 785 nm short-pass filter filters out the fundamental light. The THz pump intensity is tuned by a pair of THz polarizers. Both the pump and probe light beams are polarized horizontally, i.e. parallel to the optical table surface. The relative timing between the THz pump and the supercontinuum probe pulse is controlled via a translational stage. The continuum probe-pulse is focused roughly at the sample position to achieve an optimal spatial-overlap with the focused THz pump-pulse and re-collimated after the sample. The MAPI sample is kept in a Helium cryostat (OptistatCF-V from Oxford Instruments). We spectrally resolve the probe pulses by dispersing them onto a high-speed CMOS camera. The obtained transient spectra over time are further analyzed to improve the signal-to-noise ratio (Supplementary text 1.2, Fig. S8).

\subsection{Calculation of Optical Bloch Equations}

To assess the degree and dynamics of the intense THz-induced electronic interband tunneling process, we consider a two-level electronic system and numerically solve the optical Bloch equations,
\begin{equation}
\begin{split}
    \dot{\rho}_{00} & =-\frac{i}{\hbar}(V_{01}\rho_{10}-\rho_{01}V_{01}^*)+\gamma_{1}\rho_{11} \\
    \dot{\rho}_{01} & =\frac{d}{dt}\rho_{10}^*=-\frac{i}{\hbar}[\rho_{01}\Delta+V_{01}(\rho_{11}-\rho_{00})]-\gamma_{2}\rho_{01} \\
    \dot{\rho}_{11} &=-\frac{i}{\hbar}(-V_{01}\rho_{10}+\rho_{01}V_{01}^*)-\gamma_{1}\rho_{11}
\end{split}
\end{equation}

Here, the $\rho_{nm}$ is $(n,m)$th matrix element of the density operator, $\rho=\sum_k w_k|\psi_k\rangle \langle\psi_k|$, with arbitrary electronic states, $|\psi_k\rangle$, and $\gamma_{1} (\gamma_{2})$ represents the electronic relaxation (dephasing) time, respectively. The $(n,m)$th matrix element of dipole transition is $V_{nm}=-\mu_{01}\cdot E_{\mathrm{THz}}(t)$, where $\mu_{01}$ represents the transition dipole moment of the interband transition at the $\Gamma$ point of MAPI. $E_{\mathrm{THz}}(t)$ is the electric field of the THz pump pulse where we used the experimental temporal profile (Fig. 2(b)). $\Delta$ is the detuning of the incident field frequency compared to the interband transition energy. We assume that the initial state of the system is its ground state. It is noteworthy to mention that, under all experimental conditions, the peak electric field of the THz pulse is below the threshold for carrier multiplication \cite{kuehn2010coherent, hirori2011extraordinary}. Therefore, the contribution from carrier multiplication is not considered in this simulation. As the parameters, we take the transition dipole moment, $\mu_{01}$, to be 4.418 Debye\cite{berghoff2021low}, the interband relaxation time, $\gamma_1$, to be 200 ns \cite{li2023probing}, and the decoherence time $\gamma_2$ to be 100 fs. 

\subsection{Experimentally Constrained Lindblad Dynamics Simulation}
To investigate the measured dynamics of an anharmonic phonon mode in MAPI driven by intense THz pulses, we numerically solve the Lindblad master equation~\cite{Lindblad1976,Gorini1976}. The reduced density matrix $\rho(t)$ of the phonon system evolves according to
\[
\dot{\rho} = -i[H,\rho] + \sum_k \left(L_k \rho L_k^\dagger - \frac{1}{2}\{L_k^\dagger L_k,\rho\} \right).
\]
Here, $H(t)=H_0-\mu\cdot E_{\rm THz}(t)$ is the time-dependent Hamiltonian including the dipole coupling to the THz field, and $\{L_k\}$ are Lindblad operators describing dissipation and decoherence channels induced by the environment~\cite{Lindblad1976,Breuer2002}. This approach captures both the coherent light--matter interaction and the incoherent relaxation processes essential for describing realistic experimental conditions~\cite{Breuer2002}.

The simulation is performed in the eigenbasis of the anharmonic Hamiltonian $H_0$, whose eigenstates satisfy $H_0 | n \rangle = E_n | n \rangle$. To construct $H_0$, we fit the DFT-derived potential-energy surface of MAPI with a Morse potential~\cite{Whalley2016MAPIAnharmonicity,Morse1929}. We approximate the potential across the phase transition by retaining the local curvature of the DFT double-well potential obtained for pseudo-cubic MAPI, and rescale only the coordinate reduced mass to reproduce the experimentally observed fundamental frequency, $\omega_0 \sim 0.96$~THz~\cite{la2016phonon}. The resulting phonon spectrum is described by the standard Morse-oscillator form, 
$H_0 = \hbar \omega_0 \left[(n + 1/2)-\chi_e(n + 1/2)^2\right]$,
with anharmonicity parameter $\chi_e=1.7\times 10^{-4}$ (Fig. S7) ~\cite{Morse1929,Herzberg1950}. The eigenstates and eigenenergies are obtained by numerical diagonalization of this anharmonic Hamiltonian in a truncated harmonic-oscillator basis with $N_{\mathrm{basis}}=80$. For the Lindblad time evolution, we further restrict the Hilbert space to the lowest $N=20$ eigenstates, which is sufficient to capture the relevant driven dynamics while preserving the essential low-energy anharmonic structure of the soft potential~\cite{Brivio2015MAPI,Whalley2016MAPIAnharmonicity}.

The dipole interaction, $-\mu \cdot E_{\rm THz}(t)$, is calculated from the experimental THz pulse electric field, $E_{\rm THz}(t)$, and the experimentally determined base transition dipole moment $\mu_{01} = 2.57 D$ (Fig. S1(b)). The transition dipole matrix elements $\mu_{nm}=\langle n | \mu | m\rangle$ are computed numerically in the anharmonic eigenbasis $|n \rangle$. The dipole operator was assumed to be linear in the dimensionless displacement, $\mu = \mu_{scale} (a+a^{\dagger})$, where $a(a^{\dagger})$ are annihilation(creation) operators in the truncated harmonic basis. The scale factor $\mu_{scale}$ was calibrated so that the calculated fundamental transition dipole reproduces the experimental value $\mu_{01}^{exp}$. This procedure incorporates anharmonicity through the eigenstates and thereby captures the corresponding modification of transition strengths and effective selection rules~\cite{Nikitin1993,Triana2020}. Three experimental conditions are simulated with the waveforms measured by the electro-optic sampling~\cite{reimann2007table} and the peak field strengths: (i) centered at 1 THz with peak field strength of 183 kV/cm, (ii) 0.8 THz with189 kV/cm, and (iii) 1 THz with 107 kV/cm. 

The dissipative dynamics are modeled through two classes of Lindblad operators: population relaxation and pure dephasing~\cite{Breuer2002,Carmichael1999}.  Adjacent-level transitions are represented by jump operators $L_{n, \downarrow} =\sqrt{\gamma_{n, \downarrow}}|n-1\rangle \langle n|$ and $L_{n,\uparrow} = \sqrt{\gamma_{n,\uparrow}} | n\rangle \langle n-1 |$, with downward and upward rates satisfying thermal detailed balance for a bosonic bath~\cite{Breuer2002,Carmichael1999}. The level-dependent transition rates are chosen phenomenologically as:
\begin{equation}
\gamma^{\rm th}_{n,\downarrow} = \frac{1}{T_1} \left( \frac{|\mu_{n,n-1}|^2}{|\mu_{01}|^2} \right)
\frac{1+n_B(\Delta E_n)}{1+n_B(\Delta E_{01})}, \qquad
\gamma^{\rm th}_{n,\uparrow} = \gamma^{\rm th}_{n,\downarrow} \frac{n_B(\Delta E_n)}{1+n_B(\Delta E_n)}.
\end{equation}
Here, we take the experimentally observed relaxation time, $T_1 = 2.5 ~\rm ps$, as the base population-relaxation timescale ~\cite{kim2017direct}. The dipole-matrix-element scaling, $\gamma_{nm}\propto |\mu_{nm}|^2$, is motivated by Fermi's golden rule, while the Bose-Einstein factor $n_B = 1/(\exp(\Delta E_n/k_{\beta}T)-1)$, evaluated at the experimental temperature $T = 100~\rm K$, enforces detailed balance in the thermal regime and ensures relaxation toward the correct equilibrium distribution at long times~\cite{Breuer2002,Carmichael1999}. Loss of coherence without population transfer is modeled by diagonal Lindblad operators $L_n^{(\phi)} = \sqrt{\gamma_{\phi}} |n\rangle \langle n|$ with $\gamma_{\phi} = 1/T_2^* = 1/3.7 ps^{-1}$, derived from the experimental linewidth (0.15 THz) via $1/T_2^*=1/T_2 – 1/(2T_1)$~\cite{la2016phonon}. In addition, we implement a field-dependent crossover
between two dissipation regimes, since under strong THz driving the system is driven far from equilibrium, standard relaxation models based on thermal detailed balance are no longer expected to provide an adequate effective description~\cite{deLaTorre2021Nonthermal,Breuer2002}. Below the threshold field of 55 kV/cm, we retain full detailed balance with rates $\gamma_{\downarrow}^{\rm th}$ and $\gamma_{\uparrow}^{\rm th}$ as described above. Above this threshold, we phenomenologically suppress the downward and upward relaxation rates by factors of 10 and 100, respectively.

From the time-evolved density matrix $\rho(t)$, we first enforce Hermiticity by symmetrization and renormalize to unit trace, and then extract the following observables: the level populations ($P_n(t) = \rho_{nn}(t)$); the mean phonon energy ($\langle H\rangle (t) = \sum_n E_n P_n(t)$); the effective participation number ($N_{\rm eff}(t) = 1/\sum_n P_n^2(t)$); the total coherence ($C_{\rm total}(t) = \sum_{n \neq m} |\rho_{nm} (t)|^2$); and the von Neumann entropy ($S_{\rm vN}(t) = -\rm Tr(\rho(t) \ln \rho(t))$)~\cite{Bell1970,Wegner1980,Baumgratz2014,NielsenChuang2000}. The distance from equilibrium ($D_{\rm trace}(t)$) is quantified by the trace distance to an equal-energy Gibbs state. Specifically, at each time step we determine the Gibbs state, $\rho_{\rm G}(t) = e^{-H_0/k_B T_{\rm match}}/Z$, whose thermal average energy matches the instantaneous energy, $(\mathrm{Tr}(H_0\rho_{\rm G}(t)) = \mathrm{Tr}(H_0\rho(t)))$~\cite{Jaynes1957,Callen1985Thermodynamics}. The nonequilibrium distance is then evaluated as $D_{\rm trace}\!\left(\rho(t),\rho_{\rm G}(t)\right)
=\frac{1}{2}\mathrm{Tr}\sqrt{\left(\rho(t)-\rho_{\rm G}(t)\right)^\dagger\left(\rho(t)-\rho_{\rm G}(t)\right)}$, which measures the distinguishability between the instantaneous state and its equal-energy thermal reference \cite{NielsenChuang2000}. The instantaneous thermodynamic entropy production rate, $\Sigma(t)$ is computed using the Spohn formalism for open quantum systems \cite{Spohn1978,Breuer2002}. The entropy-production analysis is performed with respect to the thermal reference state $\rho_{\rm th}\propto e^{-H_0/k_B T}$ at the bath temperature T = 100 K. Using the dissipative part $\mathcal{D}(\rho)$ of the Lindblad equation, we evaluated the von Neumann entropy rate $\dot{S}_{\rm vN}(t)=-\mathrm{Tr}\!\left(\mathcal{D}(\rho(t))\ln\rho(t)\right)$, the entropy flux to thermal bath, $\Phi(t)=-\mathrm{Tr}\!\left(\mathcal{D}(\rho(t))\ln\rho_{\rm th}\right)$, and the total entropy production rate
$\Sigma(t)=\dot{S}_{\rm vN}(t)-\Phi(t)
=-\mathrm{Tr}\!\left(\mathcal{D}(\rho(t))\left(\ln\rho(t)-\ln\rho_{\rm th}\right)\right)$ which quantifies the irreversible contribution to the open-system dynamics~\cite{Spohn1978,Breuer2002}.

\backmatter





\bmhead{Acknowledgements}

J.K. and H.K. acknowledge funding by the National Research Foundation (NRF) of Korea (grant number RS-2025-00522010 and 2021R1A6A1A10042944). JJG gratefully acknowledges financial support from the Alexander von Humboldt Foundation. Parts of this research were carried out at ELBE at the Helmholtz-Zentrum Dresden - Rossendorf e. V., a member of the Helmholtz Association. The research leading to this result has been partly supported by the project CALIPSOplus under grant agreement no. 730872 from the EU Framework Programme for Research and Innovation HORIZON 2020. S.K. acknowledges support from the European Cluster of Advanced Laser Light Sources (EUCALL) project, which has received funding from the European Union’s Horizon 2020 Research and Innovation Programme under grant agreement no. 654220. N.A., I.I. and S.K. acknowledge support from the European Commission’s Horizon 2020 research and innovation programme, under grant agreement no. DLV-737038 (TRANSPIRE). 






\bigskip


\bibliography{sn-bibliography}

\end{document}